\documentclass[preprint]{aastex}

\def\simgr{\,\hbox{\hbox{$ > $}\kern -0.8em \lower 1.0ex\hbox{$\sim$}}\,}
\def\simle{\,\hbox{\hbox{$ < $}\kern -0.8em \lower 1.0ex\hbox{$\sim$}}\,}
\begin{document}
\title{SDSS 1507+52: A HALO CATACLYSMIC VARIABLE?
\footnote{Based on observations obtained at the MDM Observatory, operated by
Dartmouth College, Columbia University, Ohio State University, Ohio University, and
the University of Michigan.}
}

\author{Joseph Patterson}
\affil{Department of Astronomy, Columbia University, New York, NY 10027;
    jop@astro.columbia.edu}
\author{John R. Thorstensen}
\affil{Department of Physics and Astronomy, Dartmouth College, 6127 Wilder
Laboratory, Hanover, NH 03755-3528}
\author{Christian Knigge}
\affil{School of Physics \& Astronomy, University of Southampton, Highfield, 
Southampton SO17 1BJ, United Kingdom}

\begin{abstract} 
We report a photometric and spectroscopic study of
the peculiar cataclysmic variable SDSS 1507+52.  The star shows very
deep eclipses on the 67 minute orbital period, and those eclipses are
easily separable into white-dwarf and hot-spot components.  This leads
to tight constraints on binary parameters, with $M_1=0.83(8)
M_{\odot}$, $M_2=0.057(8) M_{\odot}$, $R_1=0.0097(9) R_{\odot}$,
$R_2=0.097(4) R_{\odot}$, $q=0.069(2)$, and $i=83.18(13)$ degrees.
Such numbers suggest possible membership among the WZ Sge stars, 
a common type of dwarf
nova.  The spectroscopic behavior (strong and broad H emission,
double-peaked and showing a classic rotational disturbance during
eclipse) is also typical.  But the star's orbital period is shockingly
below the ``period minimum" of $\sim$77 min which is characteristic of
hydrogen-rich CVs; producing such a strange binary will require some
tinkering with the theory of cataclysmic-variable evolution.
The proper motion is also remarkably high for a star of its distance,
which we estimate from photometry and trigonometric parallax as 230
$\pm$ 40 pc.  This suggests a transverse velocity of 164 $\pm$ 30 km/s
- uncomfortably high if the star belongs to a Galactic-disk
population.  These difficulties with understanding its evolution and
space velocity can be solved if the star belongs to a Galactic-halo
population. 
\end{abstract}


\section*{}
\begin{quote}
\textit{``\ldots of a most insignificant smallness, and a swift
wanderer among the stars."} -- \citet{kepler1619}, \textit{Harmonices Mundi}
\end{quote}

\section{Introduction}

Among the $\sim150$ cataclysmic variables so far discovered in the
Sloan Digital Sky Survey, one of the most peculiar is the 67-minute
eclipsing binary SDSS J150722.30+523039.8 (hereafter SDSS1507). 
\citet[][hereafter S05]{szkody05} presented a light curve 
showing sharp
eclipses, and a spectrum with the characteristic absorption lines of a
DA white dwarf.  The orbital period is very peculiar, far below the
well-known ``period minimum" of $\sim77$ min for hydrogen-rich
binaries.  This is quite difficult to understand.  However, binaries
with sharply defined eclipses offer good diagnostics for measuring the
individual masses, and so we carried out an extensive photometric,
spectroscopic, and astrometric study during 2005-7.  Here we report
the results of that study.  Another study is presented by 
\citet[][hereafter L07]{littlefair07}.
   
\section{Astrometry, Distance, and Velocity}

We have accumulated a two-year series of astrometric images of
SDSS1507 with the 2.4m Hiltner telescope at MDM Observatory.  Table 1
is a journal of the astrometric observations.

The astrometric images contain some photometric information.  While
nearly all observations were in the $I$-band, we also took four $V$-band
images on three different photometric nights, in order to
calibrate the field and program star magnitudes and colors for the 
astrometric reduction.  The $VI$ image pairs, reduced using 
\citet{landolt92} standards and corrected for orbital variation, 
gave average values $V$ = 18.34 and $V-I$ = +0.23.

We also derived differential magnitudes from the $I$-band parallax
images and folded them using the eclipse ephemeris derived from the
high-speed photometry (discussed below).  The resulting light curve
appears similar to the high-speed light curve, but with coarser time 
resolution because of the relatively long (120 to 150 s) integration 
times.  This fine agreement, obtained from many short visits spread 
over two years, emphasizes the reproducibility of the light curve.

The parallax and proper motion analysis followed procedures detailed
by \citet{thorstensen03}.  On each $I$-band image we measured 57
stars, 28 of which we used to define the reference frame.  We found a
relative parallax $\pi_{\rm rel} = -0.2 \pm 1.3$ mas, where the
uncertainty is estimated from the goodness of fit.  Correcting for the
estimated mean distance of the reference frame stars, we found an
absolute parallax $\pi_{\rm abs} = +0.7$ mas.  The scatter in the
fitted parallax for stars in the field with brightness similar to
SDSS1507 is 1.9 mas, so we adopt 2.5 mas as a conservative estimate on
the external error of $\pi_{\rm abs}$. Thus we find $\pi_{\rm abs} =
0.7 \pm 2.5$ mas.  The parallax is consistent with zero; our somewhat
conservative upper limit implies a distance $d > 175$ pc with over
90\% confidence.

Given its  distance, SDSS1507 has a strikingly large proper motion:
$[\mu_X, \mu_Y] = [-149, +58]$ mas yr$^{-1}$.  A proper motion this
large is typical of disk-population stars only a few tens of parsecs
distant, so the space velocity is clearly very large.

Combining the proper motion with the parallax, interpreting the proper
motion using the assumed CV space-velocity model discussed by
\citet{thorstensen03}, and accounting for the Lutz-Kelker parallax
bias, we obtain a purely astrometric distance $d = 280 (+72,-62)$ pc
(68\% confidence).  This is much smaller than $1/\pi$ because the high
proper motion drags the estimate to smaller distances.  The velocity
assumptions (that space velocities of CVs are generally low, with a
small high-velocity tail) basically quantify the implausibility of the
huge space velocity required by high proper motion at the simple
$1/\pi$ distance.

There is an independent constraint available from the time-series
photometry (reported below).  Eclipse ingress and egress of the white
dwarf (hereafter WD) are easily seen in the light curve, and show that
a small object of magnitude $V=18.9 \pm 0.15$ reappears when the WD
does.  The photometry discussed below reveals a WD radius of $0.0097
\pm 0.0009\ \rm{R}_\odot$, and the SDSS colors, along with our
scattered $BVI$ photometry, suggests $T_{\rm eff} = 12000 \pm 1200$
K.  Assuming that the line of sight permits full hemispheric
visibility of the WD near the contact phases of the eclipse, this
implies an absolute magnitude $M_V = 12.3 \pm 0.4$.  This in turn
yields a ``white dwarf parallax" distance of $215 \pm 45$ pc.  The
$T_{\rm eff}$ estimate ($11000 \pm 500$ K) of
L07 might be superior, since it is
based on full multicolor eclipse light curves; that implies $M_V= 12.5
\pm 0.3$, and thus $190 \pm 40$ pc.  Splitting the difference yields
$202 \pm 40$ pc.  Combining this with the astrometric result, we
obtain a final estimate $d = 233(+42,-36)$ pc, which we shall
characterize as $230 \pm 40$ pc.

At this distance, the transverse velocity of SDSS1507 (corrected
to the LSR) is $164 \pm 30$ km s$^{-1}$
-- a speedy traveler through local skies!  To judge its
significance, we examined a catalogue of all CVs of known orbital
period (an expanded version of Table 1 of \citealt{patterson84}), and
calculated transverse velocities for all 354 CVs of known $P_{\rm
orb}$, distance, and proper motion.  We excluded stars known or
strongly suspected to be helium-rich (although these stars are very
few and mostly ineligible anyway, since few of them offer distance
constraints).  The distance estimates used all the various clues
available (discussed by \citealt{patterson84}), with highest weight
accorded to parallaxes and detections of the secondary.  The proper
motions came mainly from the UCAC-2 \citep{zacharias04}, or from
published studies of individual stars, or from the Dartmouth parallax
program. No correction for solar motion was applied.
 
\begin{figure}
\epsscale{0.81}
\plotone{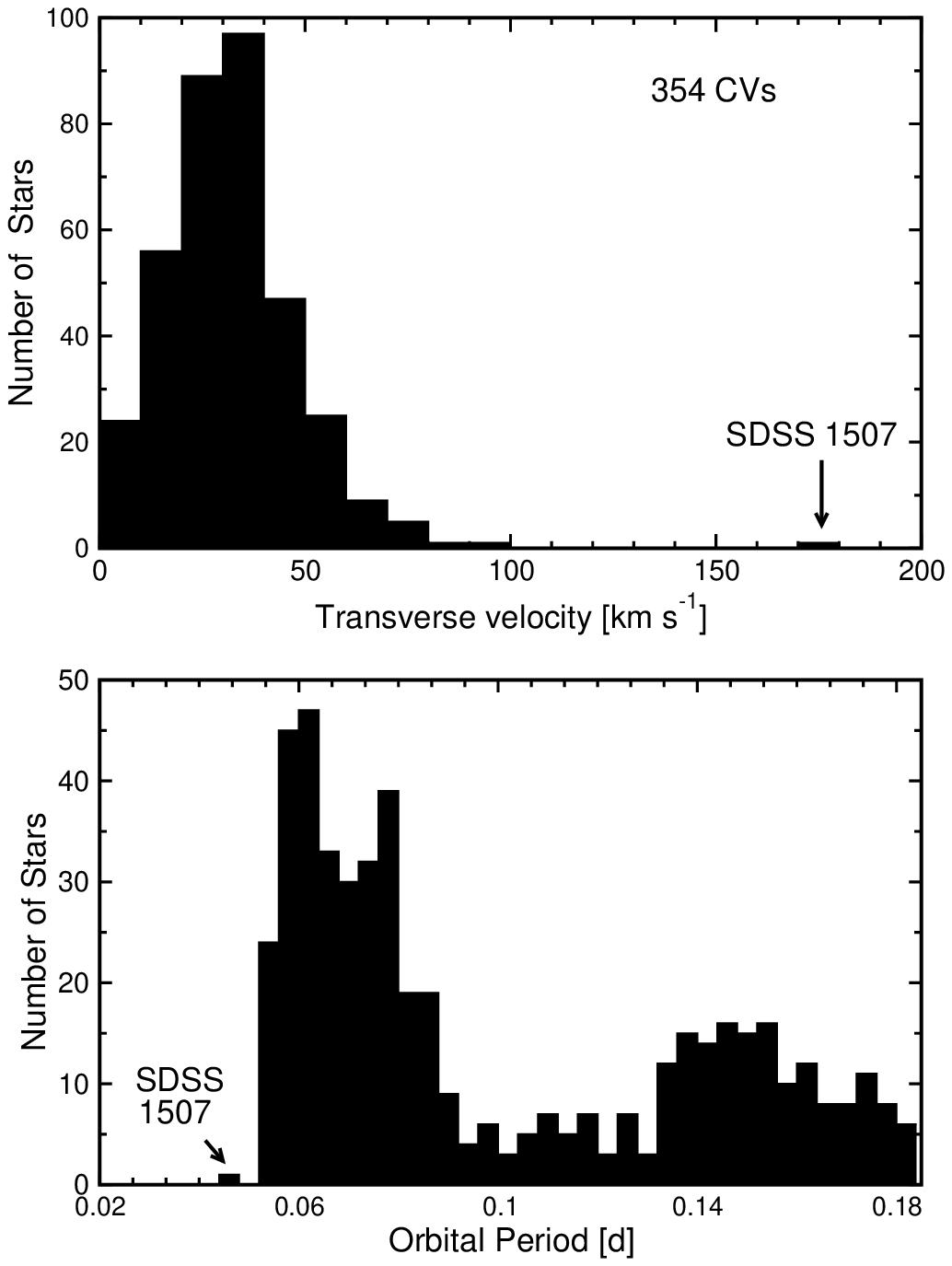}
\caption{{\it Upper frame:} Distribution of CVs in transverse velocity.
All 354 H-rich CVs with proper-motion and distance estimates are 
shown.  The average is 33 km s$^{-1}$, and the average uncertainty is $\sim$12 
km s$^{-1}$ (mostly arising from errors in distance).  SDSS1507 sits alone 
with 182 $\pm$ 30 km s$^{-1}$.  One other outlier, BF Eri, is far off-scale at 
400 km s$^{-1}$ (see text for discussion).  {\it Lower frame:} distribution of 
H-rich CVs with known orbital period.  Easily seen are the 2-3 hr period
gap, the 77 min period minimum $\ldots$ and SDSS1507, sitting by itself.
} 
\label{fig:vtrans}
\end{figure}

The upper frame of Fig.~\ref{fig:vtrans} shows the distribution of transverse
velocities.  The velocities average 33 km s$^{-1}$, a plausible value
for 1 M$_{\odot}$ stars in the thin-disk population.  There are two
outliers: SDSS1507 at 174 km s$^{-1}$ (ignoring the solar motion
correction boosts its transverse velocity slightly),
and BF Eri at $\sim 400$ km
s$^{-1}$ \citep[][not shown here, but discussed in Section 9
below]{sheets07}.  The lower frame shows the distribution with $P_{\rm
orb}$, and includes 482 stars, since the only admission credential is
a known $P_{\rm orb}$.  This illustrates the familiar 2-3 hr period
gap, the pile-up of stars at the 1.3 hr minimum period $\ldots$ and
SDSS1507, a solitary intruder sitting aloof at 1.1 hr.  These two
oddities are the motivation for this paper.

Just for the record, we can use our measurement of the emission-line
mean velocity $\gamma$ ($-46$ km s$^{-1}$, discussed below) as an
indicator of the star's systemic radial velocity; at face value this
gives a space velocity (referred to the LSR) of $167 \pm 30$ km s$^{-1}$.
Very speedy.  The components are $[U,V,W] = [+135, -93, +31]$ km s$^{-1}$.

\section{Spectroscopy}

During 2006 Jun 16-19, we obtained 53 spectra of 300 s each, with a
total exposure equivalent to 3.9 orbital periods.  We used the Hiltner
2.4 m telescope, modular spectrograph, and a thinned SITe CCD
detector, yielding 2.0 \AA\ per pixel and 3.5 \AA\ FWHM resolution in
the range 4300-7500 \AA , with vignetting near the shorter
wavelengths.  The transparency was good throughout, but the spectra
varied in quality due to seeing and guiding variations.  For
wavelength calibration we used a comparison spectrum taken during
twilight, shifted to match the zero point derived from night-sky lines
in the individual exposures; a cross-check using OH bands in the far
red showed this procedure to be accurate to typically 5 km s$^{-1}$.
For reductions we mostly followed standard IRAF-based procedures, but
in place of the IRAF {\it apsum} routines we used an original coding of
the optimal extraction algorithm published by \citet{horne86}.  We
flux-calibrated the spectra using numerous observations of standard
stars obtained during twilight on clear nights throughout the run.

\begin{figure}
\epsscale{0.9}
\plotone{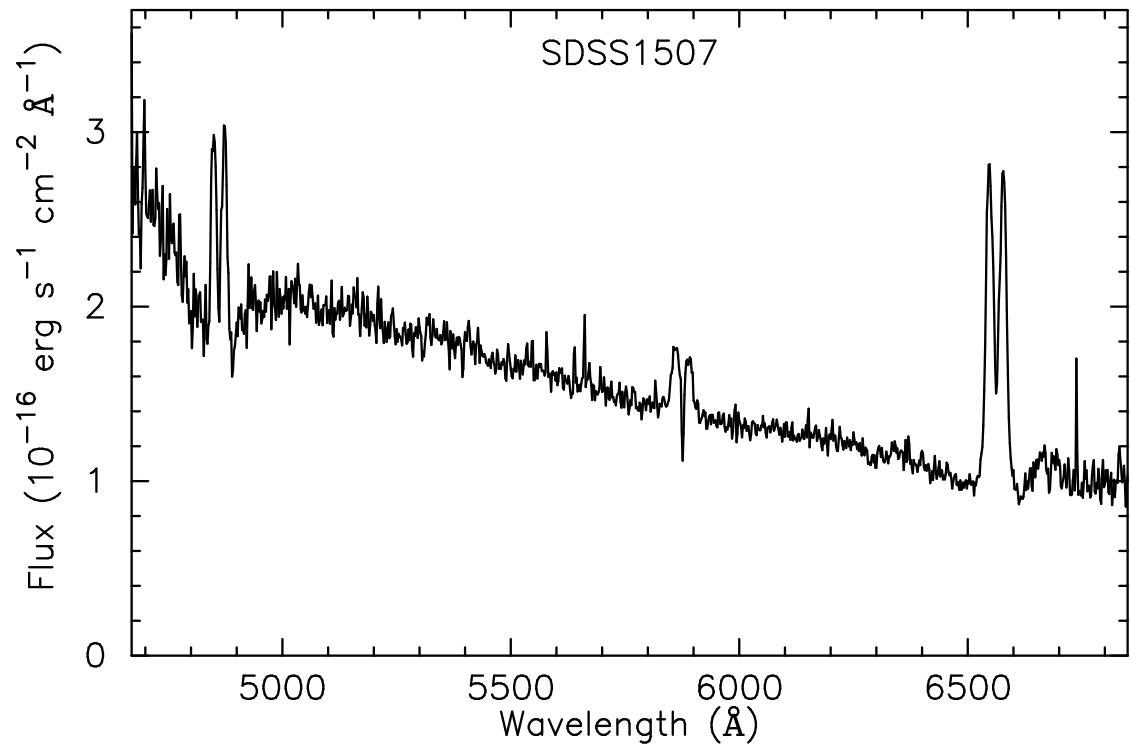}
\caption{ The average flux-calibrated spectrum, showing the broad
white-dwarf absorption, and the doubled H and He emission lines
typical of an edge-on disk.}  
\label{fig:spec}
\end{figure}

Fig.~\ref{fig:spec} shows the average flux-calibrated spectrum, and
Table 2 gives measurements of spectral features.  The blue
continuum and absorption around H$\beta$~indicates that the WD
contributes strongly to the total light.  The H$\alpha$~and H$\beta$β
emission lines are double-peaked, with the H$\alpha$~peaks
separated by 1320 km s$^{-1}$.  The continuum is noisy, because of the
star's faintness -- synthetic photometry using the \citet{bessell90}
passband gives $V = 18.4$, in good agreement with the filter
photometry.  This average spectrum is very similar to the SDSS
spectrum (S05, L07).   

\begin{figure}
\epsscale{0.9}
\plotone{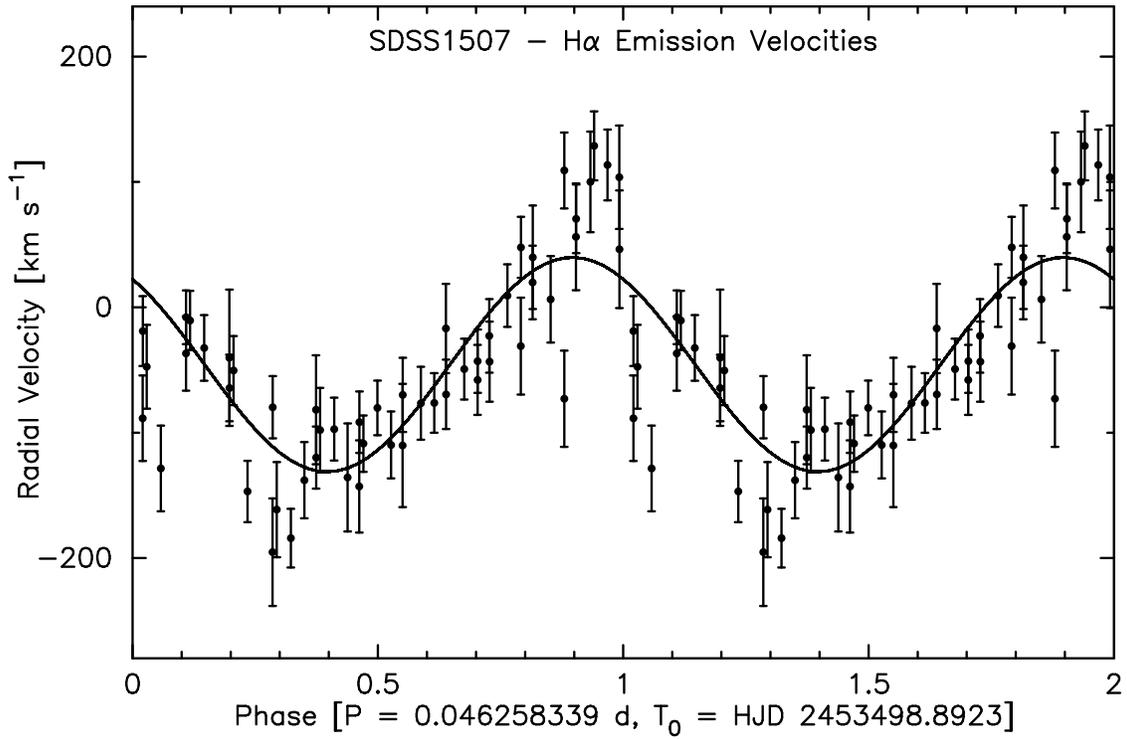}
\caption{Radial velocities of the H$\alpha$ wings, folded on the eclipse
ephemeris.  The best-fitting sinusoid is shown; it has 
semi-amplitude $K_1 = 72 \pm 14$ km s$^{-1}$, mean 
$\gamma = 46$ km s$^{-1}$, and reaches 
maximum radial-velocity at φ$\phi =0.90 \pm 0.03$.}
\label{fig:foldvel}
\end{figure}

We measured radial velocities of H$\alpha$ using a double-Gaussian
convolution method \citep{schneider80,shafter83}, with Gaussians of
370 km s$^{-1}$ FWHM separated by 2200 km s$^{-1}$.  In effect, this
measured the steep sides of the line profile.  The observations were
not distributed to optimize period finding, since the period was
already known from eclipses; even so, a period search of the
resulting time series recovers the correct orbital period.  There is
no evidence for any other period in the velocities.  Fig.~\ref{fig:foldvel} shows
the velocities folded on the eclipse ephemeris, together with the
best-fitting sinusoid. 
The velocity
reaches maximum redshift at $\phiφ=0.90 \pm 0.03$, although the WD
must be in maximum recession at $\phiφ= 0.75$.  This is very commonly
seen in cataclysmic variables; the disk's rotational velocities are
large compared to any plausible orbital motion of the WD, so a slight
periodic asymmetry in the disk's line emission can create a large
spurious radial velocity signal.  It is famously cursed as the
``phase-shift problem" (\citealt{stover81, hessman87, thorstensen91};
Sec.  2.7.6 of \citealt{warner95}).  There is now depressing evidence
that the phase-shift problem can even contaminate the ultraviolet
absorption lines, thought to arise from near the WD \citep{steeghs07}.

\begin{figure}
\epsscale{0.9}
\plotone{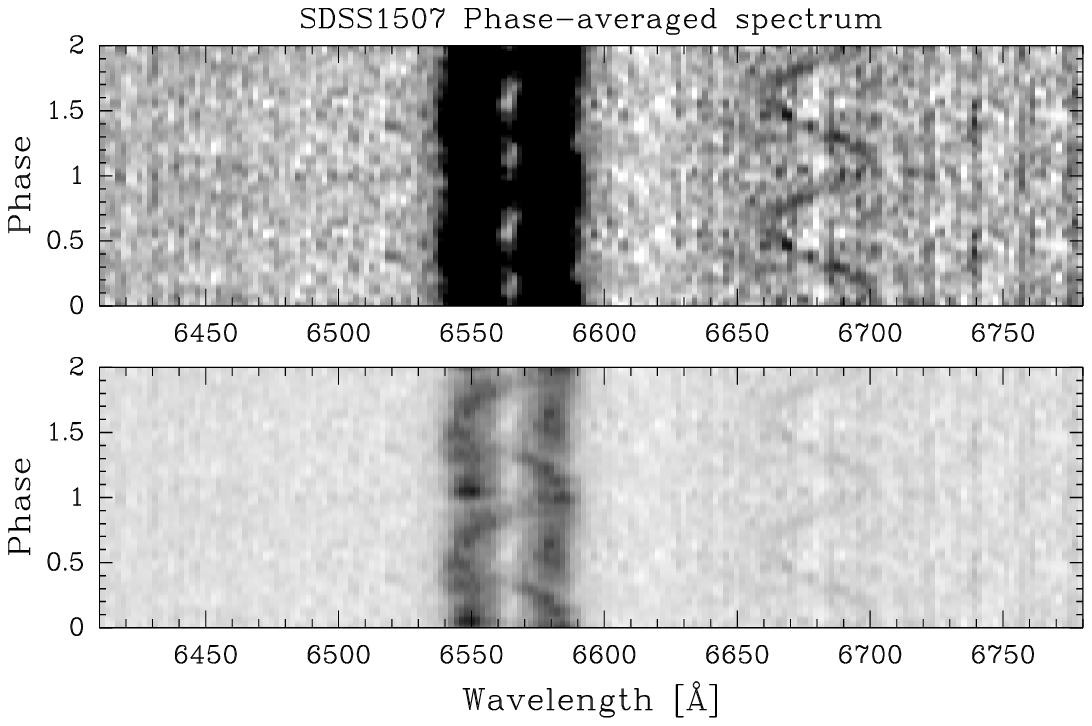}
\caption{A greyscale representation showing the changes of line
profile with orbital phase.  Both H$\alpha$αand He I $\lambda$6678 
show S-waves, as do the other useful lines (H$\beta$ 
and He I $\lambda$5876, not shown). }
\label{fig:trail}
\end{figure}

To make weaker features visible, we combined our spectra into a
two-dimensional greyscale representation using procedures explained by
\citet{taylor99}.  The individual spectra were rectified before
combining (suppressing continuum variations), and cosmic rays were
edited out by hand.  Fig.~\ref{fig:trail} shows the region around H$\alpha$ and
He I $\lambda6678$.  A prominent `S-wave' is seen in both lines, and
is especially clean in the He I line.  A similar S-wave is seen in HeI
$\lambdaλ5876$, and less prominently in H$\beta$.  We measured the
positions of these by eye in an image display; sinusoidal fits to
these measures gave nearly identical results.  The half-amplitude of
the S-wave motion is 840 $\pm$ 30 km s$^{-1}$, and the S-wave
velocities cross from blue to red at $\phi =0.86 \pm 0.01$.  

The amplitudes and phases of these radial-velocity variations merit
comparison to WZ Sge, the granddaddy of the short-period systems.  In
SDSS1507 the centroid of the broad emission lags the WD by 0.15 $\pm$
0.03 cycles, and in WZ Sge the lag is 0.12 $\pm$ 0.02 \citep{spruit98,
gilliland86}.  Their respective S-waves reach maximum radial-velocity
at $\phiφ=0.11$ and 0.13.  This suggests great similarity in the
emission-line structures.  In particular, it underlines that
phase-shift woes significantly affect both stars, and likely for the
same (unknown) reason, preventing -- as usual -- any direct
interpretation of $K_1$ as the dynamical motion of the WD.

\section{High-Speed Photometry}

\subsection{Data Taking}

Time-series differential photometry, relative to a nearby comparison
star, was obtained on 20 nights in 2005-7, with a CCD photometer on
the 1.3 m and 2.4 m telescopes of MDM Observatory.  We typically chose
an integration time of 15 s, and usually employed a ``wide blue"
filter to better define the deep eclipses, and to maximize sensitivity
to weak signals.  The observations spanned 55 binary orbits, plus a
few orbits in $B$, $V$, and $I$ filters.  Outside eclipse, the star's
mean brightness was $V = 18.3$.

\subsection{Light Curves and Eclipses}

\begin{figure}
\epsscale{0.7}
\plotone{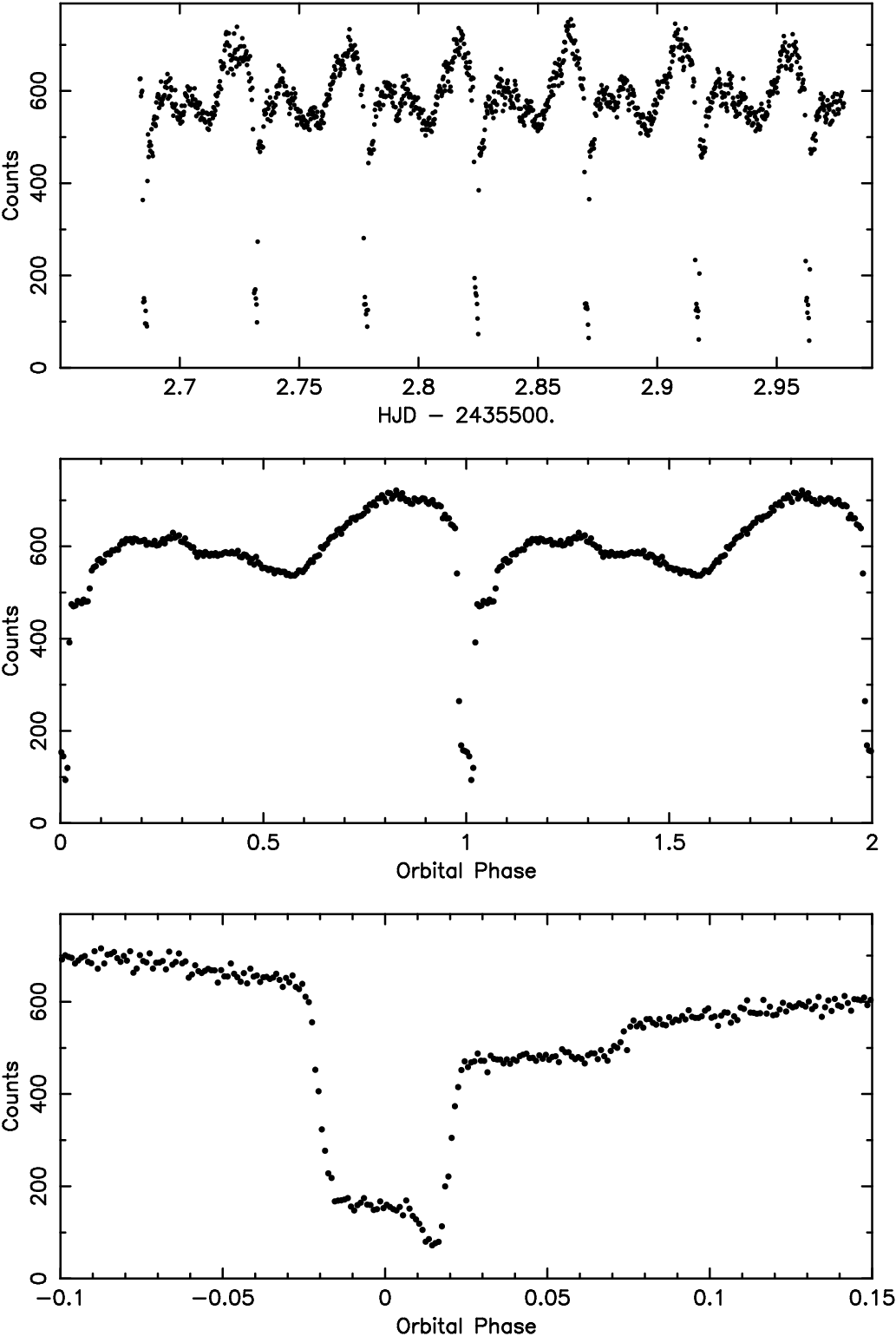}
\caption{{\it Top frame:} one night's light curve, rendered on
an intensity scale.  {\it Middle frame:} the
composite of all 2006 light curves, folded on $P_{\rm orb}$. 
{\it Bottom frame:} an expanded view of the eclipse
region, showing the double structure in the eclipse.  White-dwarf
mid-ingress and mid-egress occur at $\phiφ= \pm 0.0203$, and bright-spot mid-
ingress and mid-egress occur at $\phiφ= 0.0114(7)$ and 0.0735(6).}
\label{fig:onenight}
\end{figure}

A typical night's light curve is shown in the top frame of 
Fig.~\ref{fig:onenight}; and indeed,
every detail of the light curve repeated almost exactly on
every night of observation.  The main features are periodic humps and
sharp dips, which recur every 67 minutes.  The mean 2006 light curve
is shown in the middle frame, which has been converted to an intensity
scale, where the comparison star is arbitrarily scaled as 1000 counts.
This shows that the dip consists of two distinct parts: a WD eclipse
which is about 70\% of the mean light, and a ``bright-spot" (BS)
eclipse which is about 15\% of the mean light.  The bottom frame shows
a close-up view of the eclipse region.  This too is a mean orbital
light curve, but we achieved a higher effective by calculating
the exact orbital phase at each mid-exposure, and using
800 bins per orbit.  This was only possible because the light curve
repeats so exactly, and because we observed for many orbits.

\begin{figure}
\plotone{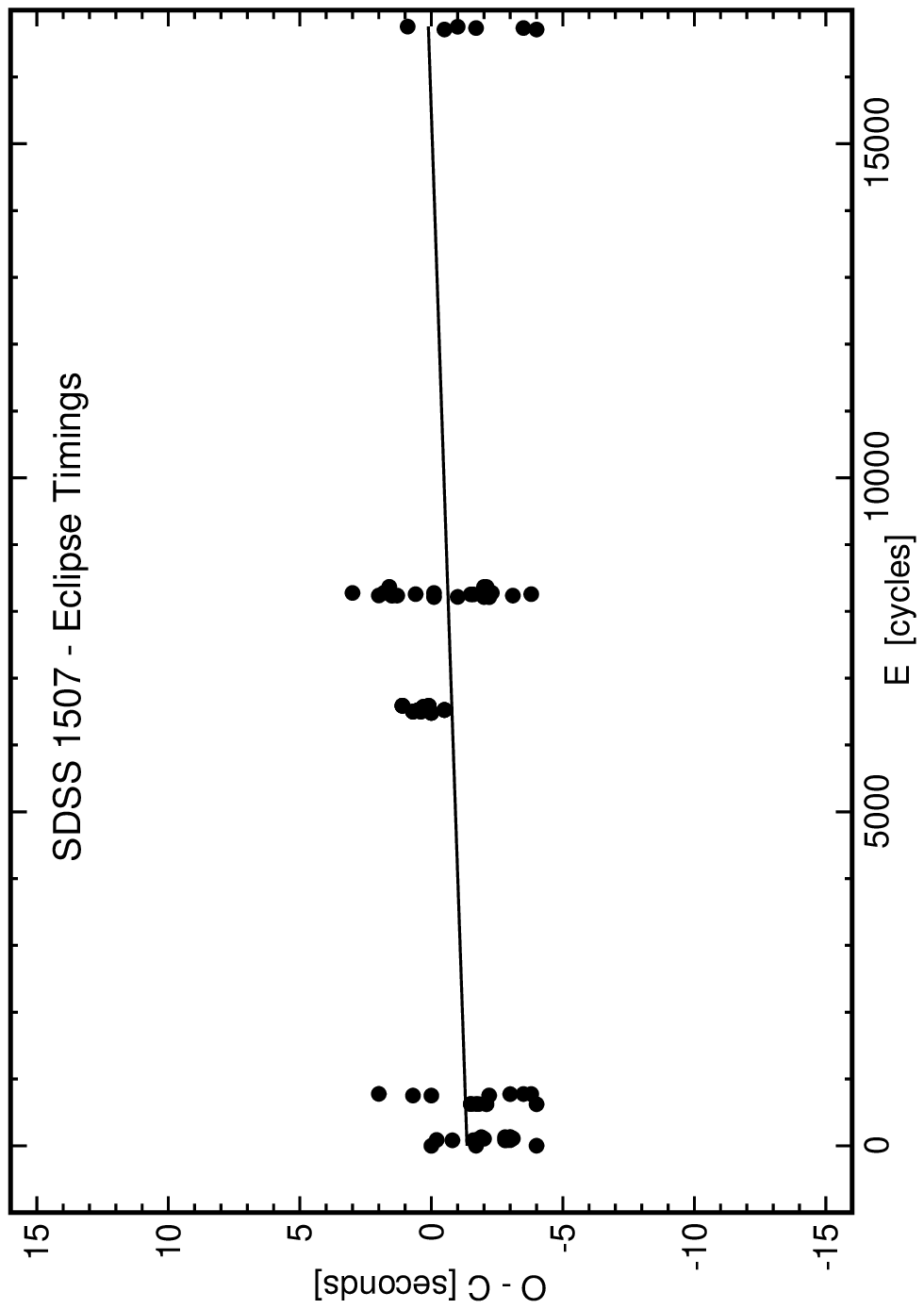}
\caption{O-C diagram of mid-eclipse timings, relative to a test
period of 0.04625834 d.  The straight line shows the best fit,
Eq. (1).}
\label{fig:ominusc}
\end{figure}

We measured mid-ingress and mid-egress times of the WD eclipse (the
sharp transitions at $\phi = \pm 0.0203$) for each eclipse, and
consider the midpoint between these precisely measurable events as the
time of mid-eclipse.  These times are presented in Table 3, and shown
in the O-C diagram of Fig.~\ref{fig:ominusc} (along with the L07 timings,
which are given triple weight since they show a lower dispersion).  
The slight upward trend indicates a
small correction to the test period, and the straight-line fit
satisfies all timings to an accuracy of 2 s.  Since the original time
series have a time resolution of $\sim$22 s (15 s integration plus 7 s
readout), this is a satisfactory fit.  Thus the best orbital
ephemeris is 
\begin{equation} 
\hbox{Mid-eclipse} = \hbox{HJD } 2453498.892264(9) + 0.0462583411(7) E.    
\end{equation}

The lesser component (BS) of the periodic eclipse is often too
weak to measure reliably in individual eclipses, but is easily measured in the
yearly means.  In 2006, mid-ingress and mid-egress occurred at $\phi
= 0.0110(10)$ and 0.0735(10),  In 2005, these events occurred at
$\phi = 0.0118(8)$ and 0.0735(8).  In 2007, a coarser measurement
gave 0.010(2) and  0.074(2).  Thus all the timings were consistent
with 0.0114(7) and 0.0735(6).

Finally, the full ingress/egress durations of the WD eclipse
were measured to be 35 $\pm$ 4 s, by fitting the yearly light curves
to a model in which a tilted knife-edge (the secondary's limb) occults
a bright sphere with limb-darkening.  

\subsection{White-Dwarf Pulsations}

We searched for other periodic signals by calculating power
from the Fourier transform of the light curve.  We did this
for each night with more than 4 hours observation, and then averaged
to find the mean nightly power spectra in 2005 and 2006, after removal
of eclipses (which severely contaminate the raw power spectra).  Most
of the significant features were merely harmonics (up to the ninth) of
the orbital frequency ω$\omega_o$; so we concluded that even apart
from the obvious eclipses, the orbital waveform is still a very strong
contaminant.  Hence we ``cleaned" the light curves by subtracting
harmonics fitted to amplitude and phase.  We then re-segregated these
yearly cleaned light curves, and calculated the nightly power spectra.

\begin{figure}
\epsscale{0.79}
\plotone{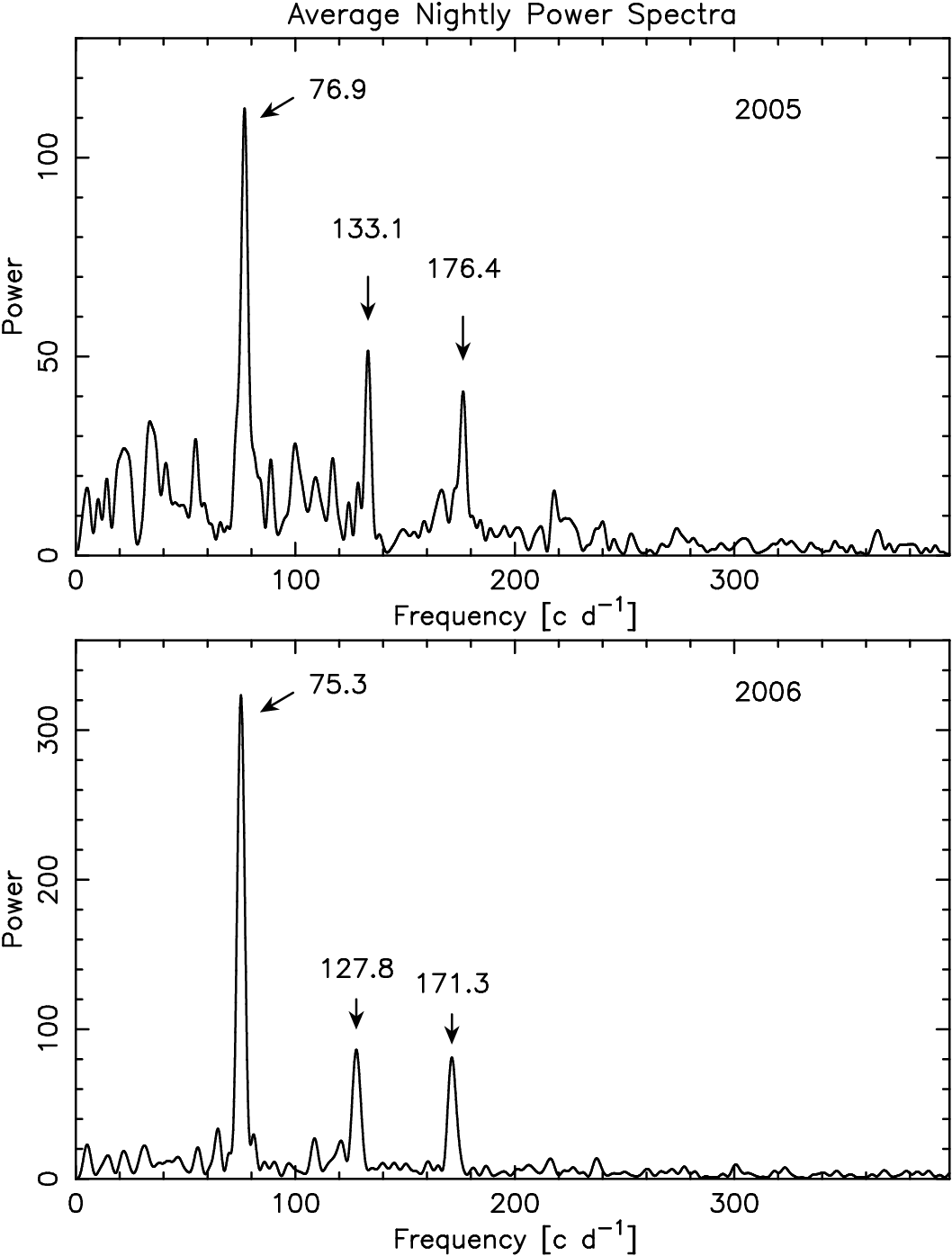}
\caption{Average nightly power spectra in 2005 and 2006; significant
features are labeled with their frequencies ($\pm$ 0.6 cycles d$^{-1}$).  The 
nightly time series have been prewhitened by removal of the 
orbital waveform.}
\label{fig:nightlypower}
\end{figure}

The yearly means are shown in Fig.~\ref{fig:nightlypower},  
with arrows flagging
significant features with their frequency in cycles per day ($\pm$
0.5).  Signals near 76, 130, and 175 cycles d$^{-1}$ appear, though with the
exact frequencies somewhat unstable.  Nothing was found in the range
400-1000 cycles d$^{-1}$.

\begin{figure}
\epsscale{0.77}
\plotone{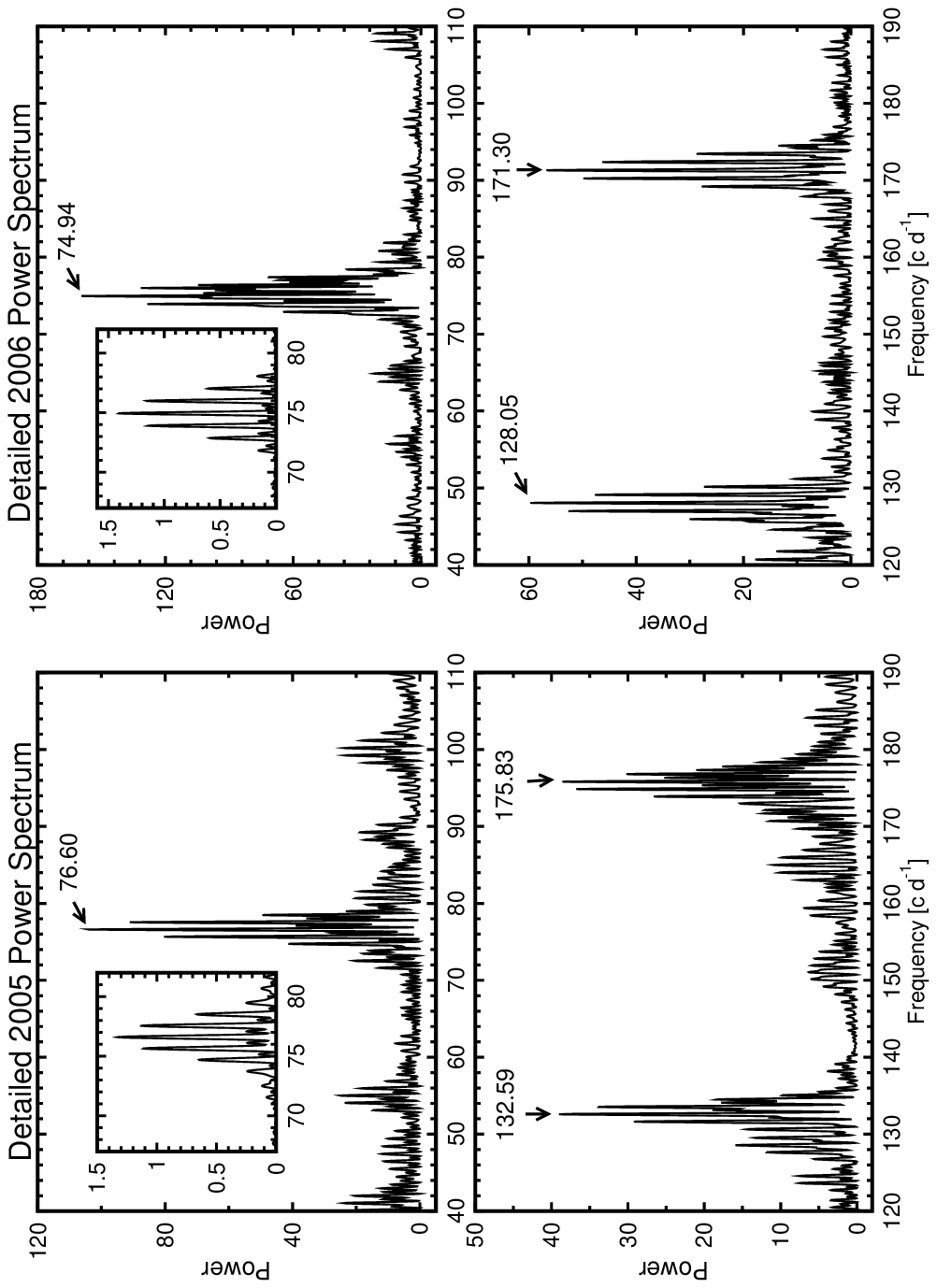}
\caption{Detailed power spectra of the several-night time series in
2005 and 2006; significant features are labeled with their 
frequencies in cycles d$^{-1}$.  Errors in 2005 and 2006, respectively, are 0.06 
and 0.04 cycles d$^{-1}$.  Inset are power spectra of an artificial signal 
sampled exactly like the actual data.  In 2006, the high-frequency   
signals are approximately consistent with a constant amplitude and 
phase, while the 75 cycles d$^{-1}$ signal is not.  In 2005, all three signals 
show an intrinsic unresolved (or undeciphered) fine structure.}
\label{fig:detailedpower}
\end{figure}

Single-night power spectra are not a good detection tool for
frequencies this low: too few cycles elapse in the course of a night.
To increase sensitivity and accuracy, we spliced consecutive nights
together (JD 2453502-4 in 2005 and JD 2453878-81 in 2006), and show
the resultant power spectra in Fig~\ref{fig:detailedpower}. The 2005 light curve showed
a primary signal at 76.60 $\pm$ 0.06 cycles d$^{-1}$, with another at
132.59 and a complex near 176 cycles d$^{-1}$.  As the picket-fence
structure in the figure suggests, each frequency has some possibility
of daily cycle count error.  The 2006 light curve showed apparent
counterparts of each signal, but with frequencies red-shifted by a few
percent: to 74.94, 128.05, and 171.30 (all $\pm$ 0.04) cycles d$^{-1}$.
The 2006 coverage was more extensive, and daily cycle count error is
very unlikely.  We assume that these trios represent basic frequencies
in the star, and will call them ω$\omega_1$, $\omega_2$, and
$\omega_3$, in order of increasing frequency.  All are incommensurate
with the orbital frequency $\omega_o$, but the weaker signals appear
to be linked to each other since $\omega_3 - \omega_2 = 2ω\omega_o$
within measurement error during both years.
   
Inset are the power spectra of artificial signals sampled exactly like
the real data; these show how a pure sinusoid at that frequency should
appear, in the absence of noise.  For a strong signal any deviation
from this window pattern signifies an underlying complexity of the
signal.  The strong signal at 74.94 cycles d$^{-1}$ in 2006 shows
significant excess power in its blue wing, which basically attests to
an unresolved fine structure in the signal.  The same is not evident
for the signals at higher frequency.  In 2005, the strongest signal
again is complex; that is probably also true for the weaker signals,
but is not quite provable amid the surrounding noise.
   
In summary, SDSS1507 shows a dominant signal near 1140 s, with at
least one additional component at slightly shorter period.  When we
attempted to parse the underlying fine structure, we could not do so
unambiguously.  But the components must be variable in frequency
and/or amplitude, and there are probably at least three of them (many
trials with only two components failed to give a satisfactory fit).
Weaker signals near 660 and 500 s are also evident; their frequencies
appear to be consistently separated by exactly $2ω\omega_o$.

The light curve and spectrum show that a WD dominates the light, so
these signals are very likely to be seated in the WD.  The
complexities observed here (multiple and incommensurate periods, with
fine structure) are often found in the nonradial pulsations of WDs,
the ZZ Ceti stars.  So SDSS1507 has joined the small club of CVs with
nonradial WD pulsations, which we shall call the ``GW Lib stars",
after its first and most famous member (for other members and a
review, see \citealt{mukadam07}).

\section{Eclipse Analysis and the Binary Parameters}

The timings of the WD and BS contact points can be used, essentially
following \citet{wood89}, to estimate the binary parameters of
SDSS1507.  These estimates are listed in Table 4 and have been
determined via the following steps.

\begin{figure}
\plotone{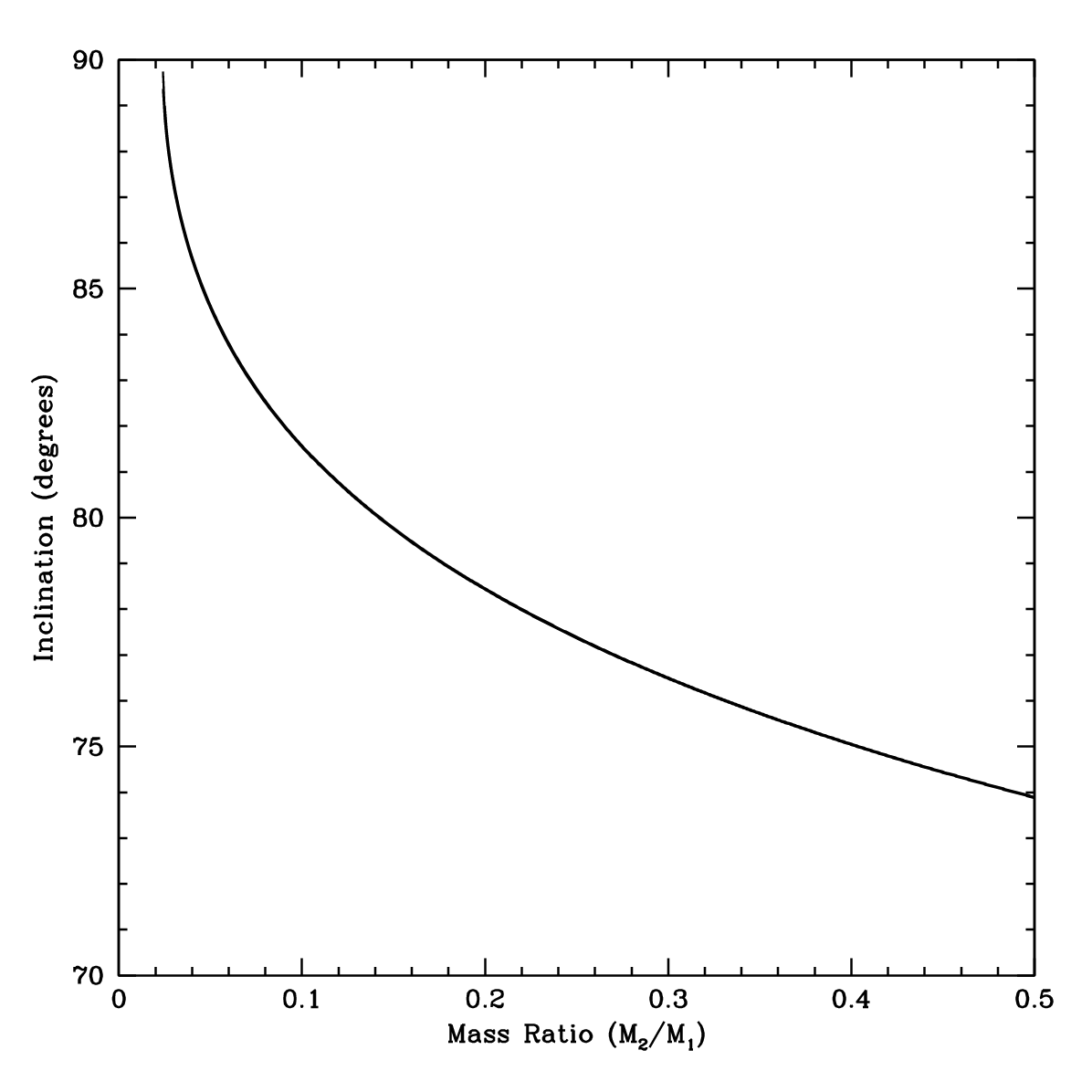}
\caption{The $q(i)$ dependence for φ$\Delta \phi_{\rm WD} = 0.0406$, 
with a point-source eclipsed by a Roche-lobe filling secondary.  
Since we use {\it mid}-eclipse, this does not depend on the white 
dwarf's radius or limb-darkening.}
\label{fig:qofi}
\end{figure}

First, in a binary with a Roche-lobe-filling secondary, the full width
at
half-intensity of the WD eclipse ($\Delta \phi_{\rmφWD}$) defines the
acceptable $q(i)$ solutions.  
Our measured $\Delta \phi_{\rm WD}$ and the method
described by \citet{chanan76} implies the $q(i)$ constraint seen in
Fig.~\ref{fig:qofi}.

\begin{figure}
\plotone{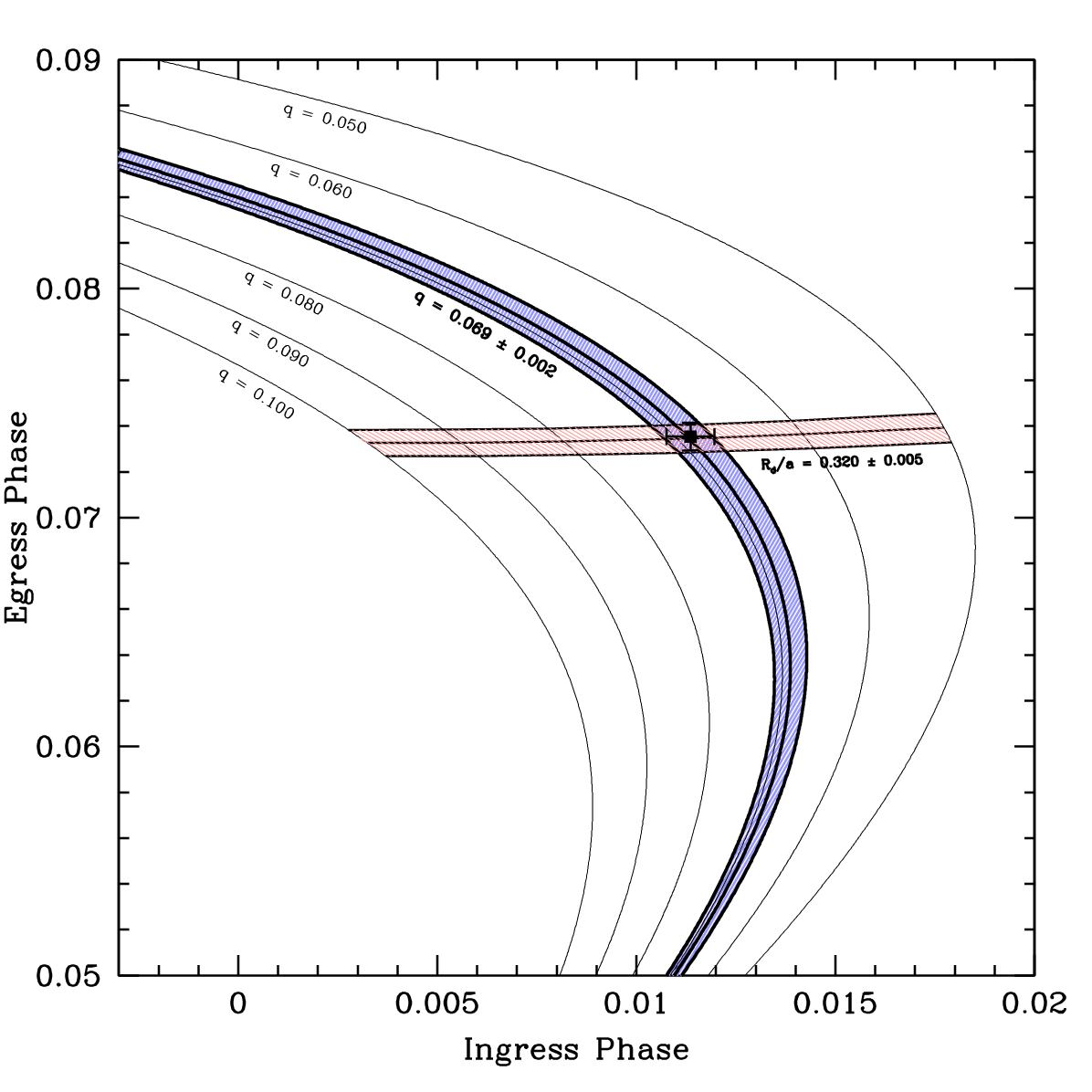}
\caption{The curves show predicted phases for mid-ingress and mid-
egress of the bright spot, as a function of $q$.  The observed phases
[0.0114(6) and 0.0735(6)] select $q=0.069(3)$.  Heavy shading shows the
curve associated with this $q$, and light shading shows the allowed 
range of disk radius.}
\label{fig:ingresspred}
\end{figure}

Second, the mid-ingress and mid-egress phases of the BS eclipse
($\phi_{\rm BSI}$ and φ$\phi_{\rm BSE}$) can be used to estimate $q$,
under the assumption that the BS lies in the orbital plane and along
the ballistic trajectory of the accretion stream from the L1 point.
We have therefore calculated stream trajectories for a range of mass
ratios \citep{flannery75} and worked out the ingress/egress phases of
a set of points along each trajectory using methods similar to that
described by \citet{chanan76}.  Fig.~\ref{fig:ingresspred} shows the calculated stream
trajectories for various mass ratios in the ($\phi_{\rm BSI},
\phi_{\rmφBSE}$) plane, along with the best measurement of those
phases.  We thus estimate $q=0.069(2)$, and Fig.~\ref{fig:qofi} then implies
$i=83.18(13)$ degrees.

Third, if the BS photometric center lies at the disk's outer edge,
then this method also provides an estimate of the disk's radius.
Fig.~\ref{fig:ingresspred} yields $R_{\rm disk}/a=0.320(5)$.

\begin{figure}
\plotone{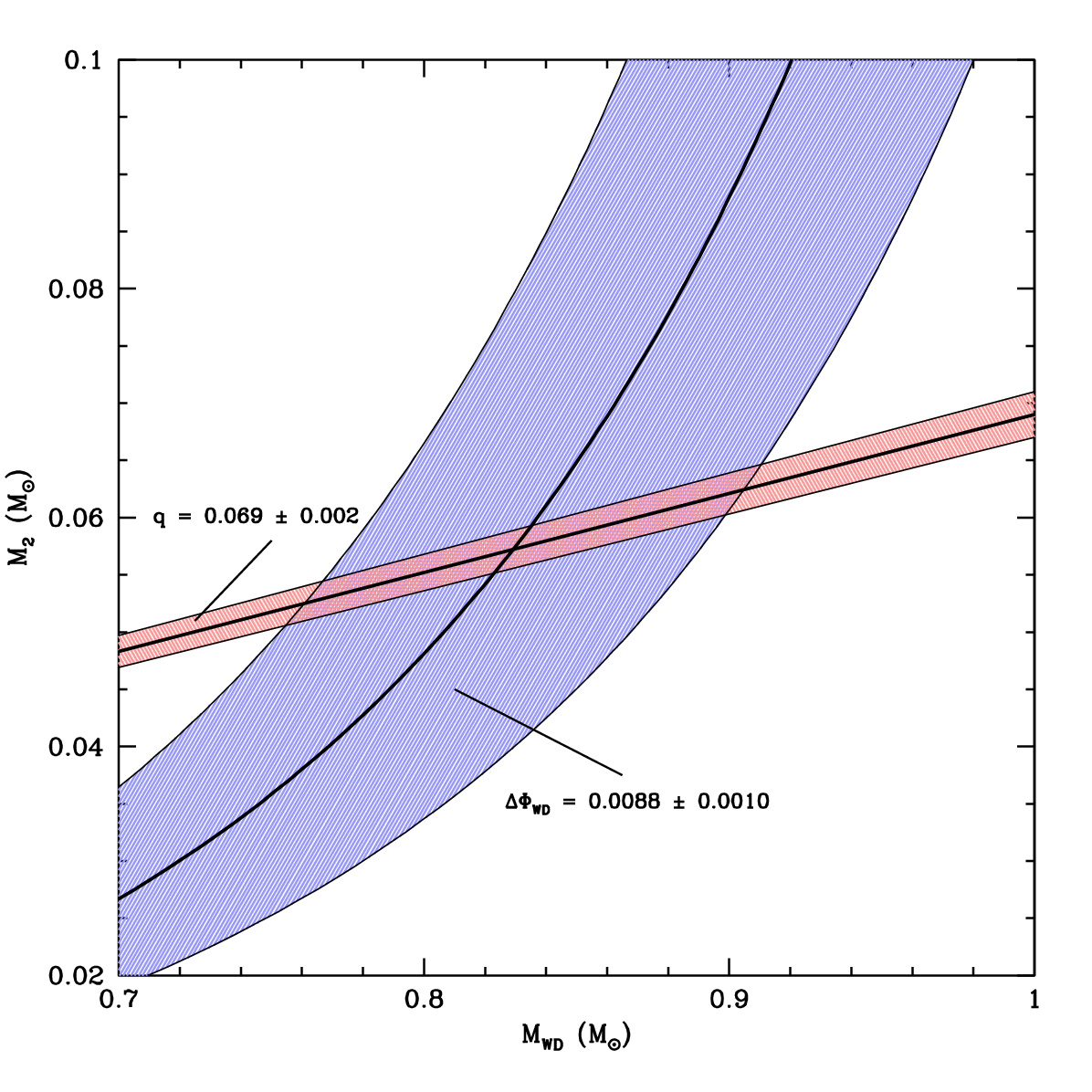}
\caption{The shaded regions express the constraints on $q$ (0.069 $\pm$ 
0.002) and the ingress/egress time of the white dwarf (35 $\pm$ 4 s).  
The star should live in the black region, with $M_2 =0.057(8)$  
M$_{\odot}$, $M_1=0.83(8)$ M$_{\odot}$.}
\label{fig:qMconstraints}
\end{figure}

Fourth, the duration of the WD ingress and egress ($35 \pm 4$ s)
is directly related to the visible diameter of the WD (e.g.
\citealt{araujo03}).  Assuming that the WD is fully visible (e.g. not
partially blocked by the disk), we can use this to estimate a WD
radius for each allowed $q(i)$ pair.  Finally, we use a theoretical WD
mass-radius relation \citep{nauenberg72} to deduce a WD mass (and
therefore also $M_2$, since $q$ is known).  The situation is shown
graphically in Fig.~\ref{fig:qMconstraints}.  The best estimates are $M_1=0.83(8),
M_2=0.057(8)$ M$_{\odot}$.

The binary separation $a$ then follows from Kepler's Third Law:
$a=0.53(1)\ R_{\odot}$.  The secondary radius $R_2$ follows by
combining the $R_2/a$ relation given by \citet{eggleton83} with the
measured values for $q$ and $a$.  The result is $R_2 = 0.097(4)
R_{\odot}$.

\section{Nature and Evolution}

The component masses of SDSS1507 are reminiscent of dwarf novae
with low accretion rate and long recurrence time, sometimes called the
``WZ Sagittae stars".  Most such stars do not eclipse and thus do not
have precisely known parameters, but WZ Sge itself appears to be
similar, with likely masses near $M_2 \sim 0.05$ M$_{\odot}$, $M_1 \sim
1$ M$_{\odot}$ (e.g. \citealt{patterson98, skidmore00}) 
\footnote{A quite different value of $M_2$ (near 0.10 M$_{\odot}$) appears
commonly in the literature.  This comes from applying Kepler's Laws
with the observed $K_1$ of the emission lines.  But because these
lines are significantly phase-shifted, because the distortion
consistently increases the derived $M_2$ (see Figure 10 and Appendix A
of \citealt{patterson05-2155}), and because photometric evidence
suggests a much lower $M_2$, we favor the low-$M_2$ solution.}. Most
other class members are probably similar, on evidentiary grounds of
low $q$ and/or low accretion rate (see Table 3 of
\citealt{patterson05-2155}).  In SDSS1507 the eclipse suggests
$V=20.6(3)$ for the BS near its eclipse phase \footnote{We use this as
a fiducial because it is well-specified by the eclipse, and because it
roughly agrees with the height of the orbital hump.  But the observed
hump is merely the asymmetrical part of the bright-spot luminosity; a
slightly larger fraction of the total light could arise from the
bright spot.}; at 230 pc, this implies $M_V = 13.8(7)$, which is very
faint for dwarf novae generally.  For the total accretion light
averaged over orbital phase, we similarly have $V=19.3(3)$ and
$M_V=12.3(6)$.

Since WZ Sge itself appears to have a similar structure, and has 
a precisely known
distance (43 pc; \citealt{thorstensen03, harrison04}), we can use it
to scale estimates of SDSS1507.  The BS eclipse in WZ Sge is 11(1)\%
of the total light, which implies $V = 17.6(3)$ at the eclipse phase,
and therefore $M_V=14.4(3)$.  This substantially agrees with the
estimate of $M_V = 13.8(7)$ in SDSS1507.  The stars also resemble each
other in spectrum (strong and broad doubled emission lines, plus WD
absorption features).

      However, SDSS1507 is certainly very unusual in two respects: the high
space velocity and the short orbital period (Fig.~\ref{fig:vtrans}).  The high space
velocity is hard to explain: Galactic-disk stars of $\sim$ 1 M$_{\odot}$ 
normally show space
velocities \simle 50 km s$^{-1}$, and our estimate of 167 km s$^{-1}$ is more
suggestive of a Galactic-halo membership.  Then there is the 67 min orbital
period, a full 15\% below the 77 min period minimum.  Roche-lobe geometry
constrains CV secondaries to obey a $P_{\rm orb}√\sqrt{\rho_2}$ = constant 
relation \citep{faulkner72}, so another way
of putting this is ``an oddly short period implies an oddly dense secondary".
What would cause that?

Well, there are a few CVs with oddly short periods, but all have known or
strongly suspected high helium abundance.  This is obviously true for the
pure-helium secondaries in the AM CVn class \citep{faulkner72,nelemans01}, and
is strongly favored for a few other dwarf novae as well (V485 Cen: 
\citealt{augusteijn96}; 
EI Psc: \citealt{thorstensen02, skillman02}), on grounds
of very strong He I emission from the disk.  This does not much resemble
SDSS1507, which has helium lines of ordinary strength for low-$\dot M$ dwarf
novae (with He I $\lambda6678/ H\alpha$ 
near 0.1).  Although high He abundance is a simple and
effective way to explain a dense secondary, it does not seem to be supported
here by evidence.

Another way, strangely enough, is a metal-poor secondary. Cool
metal-poor stars lack opacity sources in the envelope; so for a given
mass, they radiate their luminosity at a smaller radius.  For late M
stars, the radii of Population II models are about 20\% smaller than
their Population I counterparts of equal mass \citep{dantona87}.  With
the $P_{\rm orb} \sqrt{\rho_2}$~relation, this moves the minimum period from 
77 min to roughly 58 min, and hence leaves plenty of room for SDSS1507.  Very
similar results were found in the models of \citealt{stehle97} 
(see their Figure 1 and Table 2).

Population II membership also explains the high space velocity
in a natural way, since solar-neighborhood members of that class
always have high velocity.  That bodes well for a scientific theory --
a single hypothesis explaining the two outstanding anomalies
({\it hypotheses non multiplicanda sine necesitate}; \citealt{ockham1330}).  But
is its a priori likelihood too low?

      Well, roughly 0.5\% of stars in the solar neighborhood are
Population II stars ``just passing through" (\citealt{gould98}) \footnote{
This percentage varies from 0.2 at the Galactic plane to 2 at $z=1$ kpc;
we estimate at $z=200 pc$, typical for CVs.}.
About 200 CVs of short orbital period are known, so finding
one Greek among the Trojans seems plausible, assuming that different
stellar populations are comparably talented in forming CVs.

     Does this prove halo membership?  Nope.  Among stars generally,
space velocities near 188 km s$^{-1}$ also include members of intermediate
populations (e.g. ``thick disk").  We do not yet have enough information
(and enough space velocity) to certify halo membership.  Future
spectroscopic study might change this, if the binary components'
pollution of each other's surface is not too problematic.

\section{Pulsations Revisited}

      Isolated WDs with nonradial pulsation - the ZZ Ceti stars - have
well-measured temperatures and masses.   Spectroscopy and colors,
along with measured parallax, accurately establish $\log g$ and $T_{\rm eff}$, and the
pulsations appear to afflict all WDs in a narrow temperature range.  For
garden-variety WDs of 0.6 M$_{\odot}$, that range appears to be 11000-12300 
K (the
instability strip).  However, theory suggests that the $T_{\rm eff}$ boundaries depend
slightly on mass.   \citet{fontaine03} calculate that the center and width
of the strip should increase with mass, and observations appear to support
this (see Figure 6 of \citealt{gianninas05}).  In particular,
the expected instability strip for the the 0.83 $\pm$ 0.08 M$_{\odot}$ WD mass in
SDSS1507 is 11400-12900 K.  Adding uncertainty for the general dearth of WD
variability studies for masses this high, we would characterize the ``expected"
range as 11300-13000 K (in the scale used by \citealt{gianninas05}; this might
vary slightly from author to author).

Fussing over these numbers is worthwhile, because we still do not know
(and neither does anybody else) whether the WDs in GW Lib binaries are
{\it merely} ZZ Cet stars, or whether accretion and/or
membership in a close binary plays an important role in the excitation and
properties of the pulsations.  On the observational side, we need to know the
WD temperatures, and available estimates are still crude.   Temperature
measures in ZZ Cet stars are typically  $\pm$ 200 K, when a fit to the full
spectrum can be made with no needed corrections.  For CVs, the uncertainty is
much larger.  Other light sources (disk plus BS) always contaminate, and this
is very difficult to subtract since there are no cases where it can be
measured in the absence of the WD.  Also, the distances to most CVs are poorly known,
so observers commonly fit only $T_{\rm eff}$ and merely assume a radius (in particular,
assume $\log g = 8$, corresponding to $M_1 = 0.6 M_{\odot}$).  Finally, most fits are fairly
poor, and are greatly improved by adding arbitrary second components of
different area and different temperature --– a little {\it deus ex machina} which
suggests that the model is somewhat flawed.

But the situation is not all bad.  Happily, the temperatures are
roughly in the range 10000-16000 K, where Lyman-$\alpha$ absorption is dominant and
temperature-sensitive.  Present constraints are summarized by 
\citet{szkody02, szkody07}:
temperatures are known for five GW Lib stars; four are near
15000 K, and one is 10600 K.  The reporting papers do not contain much
discussion of errors, but our impression is that they are likely to be at
least  $\pm$ 1400 K.  Unfortunately, none of the five have WD masses usefully
constrained by other data (and this alone implies an error of $\pm$ 1000 K,
since $\log g$ and $T_{\rm eff}$ are highly correlated in absorption-line fitting).  
That being so, and with the oddball at 10600 K (V455 And; \citealt{araujo05})
the important $T_{\rm eff}$ questions for GW Lib stars appear to be still
unanswered: do they have an instability strip, and does it agree with the ZZ
Ceti strip?

SDSS1507 is of interest here, because its WD has a well-measured mass
and radius ($\log g= 8.4$).  The star has $T_{\rm eff} =11500 \pm$ 700 K (splitting
the difference between the estimates in Sec. 2), and the ZZ Ceti
instability strip for this mass is 11300-13000 K.  So for this star,
agreement with the ZZ Ceti strip is possible.  On the other hand, the
star accretes matter and may have a metal-poor composition, so the 
``expected" instability strip is certainly an open question.
\citet{arras06} show that the theoretical 
instability strip is greatly affected by the WD mass and the composition 
of the envelope.  This challenges obervers to find ways of estimating 
the mass of these pulsating WDs.  (And, to spice up the wish list, the 
composition of accreted gas!)

\section{A Recently Formed CV?}

As this paper was being written, we received a preprint from L07
containing a similar discussion of SDSS1507.  Comparison of the two
papers might be useful.

The quality of the orbital light curves appears to be similar, and
comparison of the light curves (their Figure 1 and our Figure~\ref{fig:onenight}) shows
that the star was in a similar state (quiescence).  L07 present
filtered photometry, however, and explicitly fit a WD flux
distribution; this is likely to yield more accurate WD properties,
so their estimate of $T_{\rm eff}$ (11000 $\pm$ 500 K versus 12000 $\pm$
1200) is perhaps better.  We obtained much more photometric
coverage (55 orbits versus $\sim$6), which perhaps explains why we found
the WD pulsations.       

The accuracy of the derived binary parameters (our Table 4 and their
Table 4) is worth some attention.  There is generally good agreement
within our uncertainties.  But L07 report very small errors,
about six times smaller than ours.  In part, this is probably
because their data are slightly better (bigger telescope,
faster time resolution, all multicolor).  In part it is probably
because their method of ``parameter estimation" tends to use all the
information in the light curve to constrain each component, whereas
our method is more piece-by-piece.  Both methods are vulnerable to
their untested assumptions: e.g., that the ``white dwarf" is
actually round and fully visible; that the bright spot's photometric
center occurs at the disk periphery; that the bright spot's
photometric center occurs at the stream/disk impact point; and that
the disk is circular.  And this litany of potential vulnerabilities
is not complete; it comes from about two minutes of passing pessimism.

L07 also favor a very different model for SDSS1507: a CV recently
formed from a detached WD - brown dwarf pair (see \citealt{politano04}
for a discussion of the theory).  The idea is that the "zero-age€
assumption enables the brown dwarf to keep a relatively high density,
and not distend due to the loss of thermal equilibrium that
appears to afflict other CVs.  This formally does the trick -- oddly,
by invoking great youth rather than great age!  But we disfavor this
idea.  First, it leaves the high space velocity entirely unexplained.
Second, it requires an auxiliary postulate: that SDSS1507 is {\it unique}.
\citet{politano04} suggests that such CVs might be fairly common, because
brown dwarfs have long thermal timescales and can remain at their
natural radius for 1-2 Gyr in a CV environment (losing mass at only
$10^{-11}$ M$_{\odot}$ yr$^{-1}$).  But this should produce a significant 
``zero-age to terminal-age" spread among the secondaries, which would be 
manifest by a large dispersion about the mean in mass-radius, and 
a quite ``soft"
period minimum.   Observations seem to reveal a tight mass-radius
relation (Figure 12 of \citealt{patterson05-epsq}, Figure 6 of 
\citealt{knigge06}), and a proverbial brick wall at $P = 77$ min (Figure 1).  

Elsewhere \citep{patterson84, patterson98, patterson01} we have speculated why the
observed period minimum is shifted from the pure-GR theoretical  value
of 70 min (because mass-transfer is slightly enhanced over the pure-GR
expectation), and why it is a brick wall (because the enhancement is
roughly the same in all CVs) \footnote{It remains
possible, and we earlier advocated \citep{patterson98}, that the period
minimum actually {\it is} soft, because the enhancement varies significantly
from star to star.  This is an important point often missed in the
literature on the subject.  It would explain the much-discussed
absence of a ``spike" at 77 min \citep{patterson98, king02}. 
But if so, then the theoretical minimum should range
from 77 to (say) 90 min - which still leaves 77 as the minimum, and a
67 min H-rich binary still unexplained and mysterious.}.
But regardless of the
merit of those speculations, a H-rich system at $P = 67$ min is indisputably
odd. The L07 model can make such a binary, but has no natural way to
make only {\it one}; the star just doesn't have any close relatives.  A halo
model better survives encounter with Ockham's Razor, because it explains
both surprising facts, not merely the period, and because it does not
require any extra novelty to explain why the star is unique (the Galactic
halo is known to exist and to contribute $\sim$0.5\% of stars to the solar
neighborhood).

\section{Other Halo CVs?}

Are there other CVs which belong to the Galactic halo?  Well, they
must be extremely rare.  There are really only three grounds for
invoking halo membership: 
\begin{enumerate}
\item{Very high space velocity (decent evidence if high enough).} 
\item{Very low metallicity (quite good evidence).}
\item{Residence in a globular cluster (excellent evidence).}
\end{enumerate}

As for (1), SDSS1507 is one of just two outliers among the 354 stars
we considered.  We also inspected a secondary list of $\sim$150 CVs with
proper motion and rougher distance constraints (based on disk
``standard candle" assumptions only).  None showed remarkably high
transverse velocity ($>$100 km s$^{-1}$).  So there are just two among
500, nearly all in the solar neighborhood (within 1 kpc).

As for (2), no CV has ever been certified to be of low metallicity.
Abundance studies in CVs are still basically beyond our reach, except
in a few cases where the stars fairly scream at us (the helium stars).

As for (3), globular clusters have been closely studied for
binary-star members, and a few dozen likely CVs have been found.
Little is known about most of them.  There is one classical nova (T
Sco in M80, \citealt{baxendell02}), and one dwarf nova with a likely 6.7 hr
orbital period (V101 in M5; \citealt{neill02}).  Since CVs in globular
clusters probably form by capture processes, these stars are in a
special category.

To these three grounds we can add a fourth of much lower scientific
weight, but by far the most common argument for invoking halo
membership: height above the galactic plane, suggested by the galactic
latitude and an assumed distance.  This has revealed some odd stars,
but just a tiny handful at heights above 1 kpc; and even in that
lonely place, there appear to be plenty of Galactic-disk stars ``just
visiting".

So halo CVs are mighty rare.  The only other candidate in the  solar
neighborhood, BF Eri, has an even higher transverse velocity  (~400 km
s$^{-1}$ at the likely 800 pc distance), but presents a puzzle since
it shows normal-looking K-star absorption features in the spectrum
\citep{sheets07}.  This could be consistent with a Population II
origin, followed by pollution of the secondary's photosphere by ejecta
from the WD $\ldots$ or with an ordinary origin, followed by a violent
disruption event (perhaps a supernova, or debris pirated from a
passing galaxy).  However, in the spirit of Ockham's Razor, the
need for such auxiliary hypotheses could also mean that we
profoundly misunderstand the true meaning of these high velocities.


\section{Summary}

\begin{enumerate}
\item{We report a photometric, astrometric, and spectroscopic study of the
peculiar CV SDSS1507, based on 2005-7 data.  The star showed deep
eclipses recurring with a 66.7 min orbital period -- shockingly lower than
the famous 77 min ``minimum period" for hydrogen-rich CVs.}
\item{The light curve repeated exactly during each orbit, showing a
double-humped wave and quite sharply defined white-dwarf and bright-
spot components.  We measured all the contact times and used them to
construct a photometric model of the system, with details in Section 5 and
Table 4.  Apart from the strange orbital period, the model resembles a
commonplace ``WZ Sge" binary, with a 0.83(8) M$_{\odot}$ white dwarf, a 
0.057(8) M$_{\odot}$
brown-dwarf secondary, and a bright-spot at the periphery of a small disk.}
\item{The spectroscopy also looks fairly ordinary for an eclipsing low-$\dot M$
dwarf nova in quiescence: bright and broad emission lines, with typical
He/H line strengths, and accompanied by a strong s-wave which wheels about
with a semi-amplitude of 840 $\pm$ 30 km s$^{-1}$.  The line wings move
with $K_1 = 75 \pm 12$ km s$^{-1}$, and reach maximum recession at $\phi =0.90
\pm 0.03$.  As usual in CVs, these values of $K_1$ and $\phi_0$φ
disqualify $K_1$ as a diagnostic of actual binary motion.}
\item{The light curves show obvious short-period pulsations, at frequencies
incommensurate with the orbital frequency.  Because the signals are fast
(500-1200 s) and fairly coherent, and because the white dwarf dominates the
light, a white-dwarf origin seems likely.  And because they are
non-commensurate, a natural explanation is nonradial pulsation.  SDSS1507
should be considered a new member of the ``GW Lib" class: intrinsically
faint CVs with multiperiodic fast signals, probably due to nonradial
white-dwarf pulsation.  This is the first accurate mass determination for a
GW Lib star.}
\item{The very high proper motion of 160 mas yr$^{-1}$ suggests a very nearby star.
But ``white-dwarf parallax" suggests a distance of 202 $\pm$ 40 pc; and the
trigonometric parallax of 0.7 $\pm$ 2.5 mas also rules out a nearby
location.  Our Bayesian estimate includes both constraints and yields 230
$\pm$ 40 pc.  This implies a transverse velocity of 164 $\pm$ 30 km
s$^{-1}$, a space velocity 167 $\pm$ 30 km s$^{-1}$, and galactic
components $[U,V,W] = [135,-93,+31]$ km s$^{-1}$.  Truly a swift wanderer
among the stars.}
\item{Our binary parameters appear to be in fair agreement with those derived
by L07 with a somewhat different method (but also mainly based on the
eclipse shape).  But we favor a different interpretation: as an ``ordinary
Population II CV".  This readily accounts for the star's two great
anomalies: the period is shockingly low because Population II secondaries
are smaller, and the space velocity is high because that is a
characteristic of Population II.}
\item{Very few Population II CVs have yet been identified outside of globular
clusters; among the H-rich stars, only two seem to have acceptable
space-velocity credentials.  That seems a plausible number, though, because
Population II represents $\sim$0.5\% of all stars in the solar neighborhood, and
approximately 500 CVs are eligible by our criteria (known proper motion and
distance).  The case for SDSS1507 seems particularly good, since the oddly
short period confers on its secondary the extra credential of an oddly
small radius (``subdwarf").  Searches for metallic lines in the disk, and in
the white dwarf, are likely to test these ideas further.}
\end{enumerate}

\acknowledgments
{\it Acknowledgments.} We have received extensive help with the observations from Jonathan
Kemp, Eve Armstrong, Jennifer Piscionere, and Michael Shulman.
Discussions with Isabelle Baraffe about binary-star evolution and the
structure of low-mass stars were also very helpful.  This research was
supported by the NSF through grants AST-0406813, AST-0307413, and 
AST-0708810.

\clearpage

\clearpage

\clearpage

\clearpage

\clearpage

\clearpage

\clearpage

\begin{deluxetable}{lrrrrr}
\tablewidth{0pt}
\tablecolumns{6}
\tablecaption{Journal of Astrometric Observations}
\tablehead{
\colhead{UT Date} & 
\colhead{$N_{\rm obs}$} & 
\colhead{HA$_{\rm start}$} &
\colhead{HA$_{\rm end}$} & 
\colhead{$p_X$\tablenotemark{a}} &
\colhead{$p_Y$} \\
\colhead{} &
\colhead{} &
\colhead{[hh:mm]} &
\colhead{[hh:mm]} &
\colhead{} &
\colhead{} \\
}
\startdata
2005 Jun 25 & 12 &  $+$0:26  &  $+$0:53  & $-0.69$ & $0.74$ \\
2006 Mar 13 &  5 &  $+$0:18  &  $+$0:36  & $0.80$ & $0.43$ \\
2006 May 20 & 14 &  $+$0:56  &  $+$2:24  & $-0.17$ & $0.95$ \\
2006 Jun 10 &  9 &  $-$0:08  &  $+$0:23  & $-0.49$ & $0.87$ \\
2006 Aug 29 &  4 &  $+$2:59  &  $+$3:13  & $-0.93$ & $-0.18$ \\
2007 Jan 28 &  8 &  $-$1:00  &  $-$0:36  & $0.93$ & $-0.27$ \\
2007 Mar 25 & 10 &  $-$0:22  &  $+$0:18  & $0.68$ & $0.59$ \\
2007 Jun 21 &  8 &  $+$0:02  &  $+$0:38  & $-0.64$ & $0.78$ \\
\enddata
\tablenotetext{a}{Parallax factors $p_X$ and $p_Y$ are the 
parallax displacements that a star at 1 pc distance would
undergo on that date, in $X$ (eastward) and $Y$ (northward).}
\end{deluxetable}

%

\begin{deluxetable}{lrcc}
\tablewidth{0pt}
\tablecolumns{4}
\tablecaption{Spectral Features in Quiescence}
\tablehead{
&
\colhead{E.W.\tablenotemark{a}} &
\colhead{Flux}  &
\colhead{FWHM} \\
\colhead{Feature} &
\colhead{(\AA )} &
\colhead{(10$^{-17}$ erg cm$^{-2}$ s$^{-1}$)} &
\colhead{(\AA)} \\
}
\startdata
            H$\beta$ & 25 & 9 & 40 \\
  HeI $\lambda 5876$ & 9 & 8: & 49 \\
           H$\alpha$ & 87 & 15 & 51 \\
  HeI $\lambda 6678$ & 12 & 1.5 & \nodata \\
\enddata
\tablenotetext{a}{Emission equivalent widths are counted as positive. 
No attempt has been made to measure the absorption around H$\beta$, or
to correct the H$\beta$ equivalent width for the underlying absorption.}
\end{deluxetable}

\begin{deluxetable}{rrr}
\tablewidth{0pt}
\tablecolumns{3}
\tablecaption{Times of Mid-Eclipse}
\tablehead{}
\startdata 
498.89228  &   498.93851  &   498.98475  \\
502.68543  &   502.73171  &   502.77795  \\
502.82422  &   502.87048  &   502.91672  \\
502.96301  &   503.84190  &   503.88814  \\
503.93441  &   504.85958  &   504.90583  \\
504.95209  &   527.66492  &   527.71120  \\
527.75746  &   527.80373  &   527.84998  \\
527.89624  &   533.67855  &   533.72482  \\
533.77104  &   534.69620  &   534.74245  \\
534.78871  &   534.83504  &   878.81200  \\
878.85828  &   878.90452  &   878.95079  \\
879.78347  &   879.87593  &   879.92224  \\
879.96851  &   880.70860  &   880.75486  \\
880.80111  &   880.84740  &   880.89360  \\
881.68008  &   881.72630  &   881.77258  \\
881.81879  &   885.84327  &   885.88953  \\
885.93583  &  1271.86909  &  1271.91539  \\
1272.84055 &  1272.88679  &  1273.81198  \\
1273.85826  \\
\enddata
\tablecomments{Times listed are the HJD of 
mid-eclipse, minus 2453000.}
\end{deluxetable}

\begin{deluxetable}{lr}
\tablewidth{0pt}
\tablecolumns{2}
\tablecaption{Model Parameters from Eclipse Fitting}
\tablehead{
\colhead{Quantity} & 
\colhead{Value\tablenotemark{a}} \\
}
\startdata 
Inclination $i$  &  83.18(13) deg \\
Mass ratio $q$  &    0.069(2)   \\
$M_1$ & 0.83(8) M$_\odot$  \\
$M_2$ & 0.057(8) M$_\odot$  \\
$R_1$ & 0.0097(9) R$_\odot$  \\
$R_2$ & 0.097(4) R$_\odot$  \\
$T_{\rm wd}$  & 11500(700) K   \\
$R_{\rm disk}/a$ & 0.320(5)    \\
Distance $d$ & 230(40) pc   \\
\enddata
\tablenotetext{a}{Uncertainties in the last digit(s) 
are given in parentheses.}
\end{deluxetable}

\begin{thebibliography}

\bibitem[Araujo-Betancor et al.(2003)]{araujo03} 
Araujo-Betancor, S., et al.\ 2003, \apj, 583, 437 

\bibitem[Araujo-Betancor et al.(2005)]{araujo05} 
Araujo-Betancor, S., et al.\ 2005, \aap, 430, 629 

\bibitem[Arras et al.(2006)]{arras06} Arras, P., Townsley, 
D.~M., \& Bildsten, L.\ 2006, \apjl, 643, L119 

\bibitem[Augusteijn et al.(1996)]{augusteijn96} Augusteijn, T., van 
der Hooft, F., de Jong, J.~A., \& van Paradijs, J.\ 1996, \aap, 311, 889 

\bibitem[Baxendell(1902)]{baxendell02} Baxendell, J.\ 1902, \aj, 
22, 127 

\bibitem[Bessell(1990)]{bessell90} Bessell, M.~S.\ 1990, \pasp, 
102, 1181 

\bibitem[Chanan et al.(1976)]{chanan76} Chanan, G.~A., 
Middleditch, J., \& Nelson, J.~E.\ 1976, \apj, 208, 512 


\bibitem[D'Antona(1987)]{dantona87} D'Antona, F.\ 1987, \apj, 
320, 653 

\bibitem[Eggleton(1983)]{eggleton83} Eggleton, P.~P.\ 1983, \apj, 
268, 368 

\bibitem[Faulkner et al.(1972)]{faulkner72} Faulkner, J., 
Flannery, B.~P., \& Warner, B.\ 1972, \apjl, 175, L79 

\bibitem[Flannery(1975)]{flannery75} Flannery, B.~P.\ 1975, \apj, 
201, 661 

\bibitem[Fontaine et al.(2003)]{fontaine03} Fontaine, G., 
Brassard, P., \& Charpinet, S.\ 2003, \apss, 284, 257 

\bibitem[Gianninas et al.(2005)]{gianninas05} Gianninas, A., 
Bergeron, P., \& Fontaine, G.\ 2005, \apj, 631, 1100 

\bibitem[Gilliland et al.(1986)]{gilliland86} Gilliland, R.~L., 
Kemper, E., \& Suntzeff, N.\ 1986, \apj, 301, 252 

\bibitem[Gould et al.(1998)]{gould98} Gould, A., Flynn, C., \& 
Bahcall, J.~N.\ 1998, \apj, 503, 798 

\bibitem[Harrison et al.(2004)]{harrison04} Harrison, T.~E., 
Johnson, J.~J., McArthur, B.~E., Benedict, G.~F., Szkody, P., Howell, 
S.~B., \& Gelino, D.~M.\ 2004, \aj, 127, 460 

\bibitem[Hessman(1987)]{hessman87} Hessman, F.~V.\ 1987, \apss, 
130, 351 

\bibitem[Horne(1986)]{horne86} Horne, K.\ 1986, \pasp, 98, 609 

\bibitem[Kepler(1619)]{kepler1619} Kepler, J.\ 1619, {\it Harmonices
Mundi}, tr. J. Field (1997); American Philosophical Society

\bibitem[King et al.(2002)]{king02} King, A.~R., Schenker, K., 
\& Hameury, J.~M.\ 2002, \mnras, 335, 513 

\bibitem[Knigge(2006)]{knigge06} Knigge, C.\ 2006, \mnras, 373, 
484 



\bibitem[Landolt(1992)]{landolt92} Landolt, A.~U.\ 1992, \aj, 
104, 340 

\bibitem[Littlefair et al.(2007)]{littlefair07} Littlefair, S.~P., 
Dhillon, V.~S., Marsh, T.~R., G{\"a}nsicke, B.~T., Baraffe, I., \& Watson, 
C.~A.\ 2007, \mnras, 381, 827 (L07)

\bibitem[Mukadam et al.(2007)]{mukadam07} Mukadam, A.~S., 
G{\"a}nsicke, B.~T., Szkody, P., Aungwerojwit, A., Howell, S.~B., Fraser, 
O.~J., \& Silvestri, N.~M.\ 2007, \apj, 667, 433 

\bibitem[Nauenberg(1972)]{nauenberg72} Nauenberg, M.\ 1972, \apj, 
175, 417 

\bibitem[Neill et al.(2002)]{neill02} Neill, J.~D., Shara, 
M.~M., Caulet, A., \& Buckley, D.~A.~H.\ 2002, \aj, 123, 3298 

\bibitem[Nelemans et al.(2001)]{nelemans01} Nelemans, G., 
Portegies Zwart, S.~F., Verbunt, F., \& Yungelson, L.~R.\ 2001, \aap, 368, 
939 

\bibitem[Ockham(1330)]{ockham1330} Ockham, W. of\ 1330, {\it De
Sacramento Altaris}, tr. by Birch, T.B. (Burlington, Iowa:
Iowa Lutheran Literary Board)

\bibitem[Patterson(1984)]{patterson84} Patterson, J.\ 1984, \apjs, 
54, 443 

\bibitem[Patterson(1998)]{patterson98} Patterson, J.\ 1998, \pasp, 
110, 1132 

\bibitem[Patterson(2001)]{patterson01} Patterson, J.\ 2001, \pasp, 
113, 736 

\bibitem[Patterson et al.(2005)]{patterson05-2155} Patterson, J., 
Thorstensen, J.~R., \& Kemp, J.\ 2005, \pasp, 117, 427 

\bibitem[Patterson et al.(2005)]{patterson05-epsq} Patterson, J., et 
al.\ 2005, \pasp, 117, 1204 

\bibitem[Politano(2004)]{politano04} Politano, M.\ 2004, \apj, 
604, 817 


\bibitem[Schneider \& Young(1980)]{schneider80} Schneider, D.~P., 
\& Young, P.\ 1980, \apj, 238, 946 

\bibitem[Shafter(1983)]{shafter83} Shafter, A.~W.\ 1983, \apj, 
267, 222 

\bibitem[Sheets et al.(2007)]{sheets07} Sheets, H.~A., 
Thorstensen, J.~R., Peters, C.~J., Kapusta, A.~B., \& Taylor, C.~J.\ 2007, 
\pasp, 119, 494 

\bibitem[Skidmore et al.(2000)]{skidmore00} Skidmore, W., Mason, 
E., Howell, S.~B., Ciardi, D.~R., Littlefair, S., \& Dhillon, V.~S.\ 2000, 
\mnras, 318, 429 

\bibitem[Skillman et al.(2002)]{skillman02} Skillman, D.~R., et 
al.\ 2002, \pasp, 114, 630 

\bibitem[Spruit \& Rutten(1998)]{spruit98} Spruit, H.~C., \& 
Rutten, R.~G.~M.\ 1998, \mnras, 299, 768 

\bibitem[Steeghs et al.(2007)]{steeghs07} Steeghs, D., Howell, 
S.~B., Knigge, C., G{\"a}nsicke, B.~T., Sion, E.~M., \& Welsh, W.~F.\ 2007, 
\apj, 667, 442 

\bibitem[Stover(1981)]{stover81} Stover, R.~J.\ 1981, \apj, 249, 
673 

\bibitem[Stehle et al.(1997)]{stehle97} Stehle, R., Kolb, U., \& 
Ritter, H.\ 1997, \aap, 320, 136 

\bibitem[Szkody et al.(2002)]{szkody02} Szkody, P., 
G{\"a}nsicke, B.~T., Howell, S.~B., \& Sion, E.~M.\ 2002, \apjl, 575, L79 

\bibitem[Szkody et al.(2005)]{szkody05} Szkody, P., et al.\ 
2005, \aj, 129, 2386 

\bibitem[Szkody et al.(2007)]{szkody07} Szkody, P., et al.\ 
2007, \apj, 658, 1188 

\bibitem[Taylor et al.(1999)]{taylor99} Taylor, C.~J., 
Thorstensen, J.~R., \& Patterson, J.\ 1999, \pasp, 111, 184 

\bibitem[Thorstensen(2003)]{thorstensen03} Thorstensen, J.~R.\ 2003, 
\aj, 126, 3017 

\bibitem[Thorstensen et al.(1991)]{thorstensen91} Thorstensen, J.~R., 
Ringwald, F.~A., Wade, R.~A., Schmidt, G.~D., \& Norsworthy, J.~E.\ 1991, 
\aj, 102, 272 

\bibitem[Thorstensen et al.(2002)]{thorstensen02} Thorstensen, J.~R., 
Fenton, W.~H., Patterson, J.~O., Kemp, J., Krajci, T., \& Baraffe, I.\ 
2002, \apjl, 567, L49 



\bibitem[Warner(1995)]{warner95} Warner, B.\ 1995, Cambridge 
Astrophysics Series, Cambridge, New York: Cambridge University Press,
|c1995


\bibitem[Wood et al.(1989)]{wood89} Wood, J.~H., Horne, K., 
Berriman, G., \& Wade, R.~A.\ 1989, \apj, 341, 974 


\bibitem[Zacharias et al.(2004)]{zacharias04} Zacharias, N., Urban, 
S.~E., Zacharias, M.~I., Wycoff, G.~L., Hall, D.~M., Monet, D.~G., \& 
Rafferty, T.~J.\ 2004, \aj, 127, 3043 

\end{thebibliography}
\end{document}